\documentclass[runningheads]{llncs}
\pdfoutput=1
\usepackage{amsmath}
\usepackage{amssymb}
\usepackage{multirow}
\usepackage{graphicx}
\usepackage{booktabs}
\usepackage{array}
\usepackage{url}

\newcolumntype{x}[1]{>{\centering\arraybackslash\hspace{0pt}}p{#1}}
\newcolumntype{R}[1]{>{\flushright\arraybackslash\hspace{0pt}}p{#1}}
\newcommand{\attimg}{A}
\newcommand{\patch}{I}

\newcommand{\fex}{\mathcal{F}}
\newcommand{\attconf}{\mathbf{a}}
\newcommand{\sparseparam}{\eta}
\newcommand{\attpatch}{J}
\newcommand{\scalenet}{\mathcal{G}}

\newcommand{\levels}{\mathcal{C}}
\newcommand{\nlevels}{N_\levels}
\newcommand{\lscores}{\hat{\mathbf{s}}}

\newcommand{\lprobs}{\hat{\mathbf{p}}}
\newcommand{\lprob}{\hat{p}}
\newcommand{\loss}[1]{\mathcal{L}_{\mathrm{#1}}}
\newcommand{\settransforms}{\mathcal{T}}
\newcommand{\normtwo}[1]{\left\lVert#1\right\rVert_{2}}
\newcommand{\normone}[1]{\left\lVert#1\right\rVert_{1}}
\newcommand{\model}[1]{\mathsf{M}_{\mathrm{#1}}}
\newcommand{\modeltotal}{\model{proposed}}
\newcommand{\modelNsparse}{\model{\lnot{}sparse}}
\newcommand{\modelNsmooth}{\model{\lnot{}smooth}}
\newcommand{\modelNequiv}{\model{\lnot{}equiv}}
\newcommand{\modelNWSI}{\model{\lnot{}WSI}}

\newcounter{daggerfootnote}
\newcommand*{\daggerfootnote}[1]{%
    \setcounter{daggerfootnote}{\value{footnote}}%
    \renewcommand*{\thefootnote}{\fnsymbol{footnote}}%
    \footnote[2]{#1}%
    \setcounter{footnote}{\value{daggerfootnote}}%
    \renewcommand*{\thefootnote}{\arabic{footnote}}%
    }

\begin{document}

\hyphenation{histo-pathological}

%
\title{Self-Supervised Nuclei Segmentation in Histopathological Images
Using Attention}
\titlerunning{Self-Supervised Nuclei Segmentation}
%
\author{Mihir Sahasrabudhe\inst{1,2} \and
Stergios Christodoulidis \inst{3} \and
Roberto Salgado\inst{4,5} \and
Stefan Michiels\inst{3,6} \and
Sherene Loi\inst{4} \and
Fabrice Andr\'{e}\inst{3} \and
Nikos Paragios\inst{1,7} \and
Maria Vakalopoulou\inst{1,2,3}
}
%
\authorrunning{Sahasrabudhe et al.}
%
\institute{Universit\'{e} Paris-Saclay, CentraleSup\'{e}lec, Math\'{e}matiques et Informatique pour la Complexit\'{e} et les Syst\`{e}mes, 91190 Gif-sur-Yvette, France.\and
Inria Saclay 91190, Gif-sur-Yvette, France \and
Institut Gustave Roussy, 94800 Villejuif, France \and
Division of Research, Peter MacCallum Cancer Centre, Melbourne, Australia \and
Department of Pathology, GZA-ZNA Hospitals, 2050 Antwerp, Belgium \and
Service de Biostatistique et d'Epid\'{e}miologie, Gustave Roussy, CESP U108, Universit\'{e} Paris‐Sud, Universit\'{e} Paris‐Saclay, Villejuif, France \and
Therapanacea, 75014 Paris, France}
%


\maketitle              
\begin{abstract}
Segmentation and accurate localization of nuclei in histopathological images is a very challenging problem, with most existing approaches adopting a supervised strategy. These methods usually rely on manual annotations that require a lot of time and effort from medical experts. In this study, we present a self-supervised approach for segmentation of nuclei for whole slide histopathology images. Our method works on the assumption that the size and texture of nuclei can determine the magnification at which a patch is extracted. We show that the identification of the magnification level for tiles can generate a preliminary self-supervision signal to locate nuclei. We further show that by appropriately constraining our model it is possible to retrieve meaningful segmentation maps as an auxiliary output to the primary magnification identification task. Our experiments show that with standard post-processing, our method can outperform other unsupervised nuclei segmentation approaches and report similar performance with supervised ones on the publicly available MoNuSeg dataset. Our code and models are available online\protect\daggerfootnote{\url{https://github.com/msahasrabudhe/miccai2020_self_sup_nuclei_seg}} to facilitate further research.  

\keywords{Pathology, Whole Slide Images, Nuclei Segmentation, Deep Learning, Self-Supervision, Attention Models}
\end{abstract}
\section{Introduction}

Histology images are the gold standard in diagnosing a considerable number of diseases including almost all types of cancer. For example, the count of nuclei on whole-slide images (WSIs) can have diagnostic significance for numerous cancerous conditions~\cite{ruan2018predictive}. The proliferation of digital pathology and high-throughput tissue imaging leads to the adoption in clinical practice of digitized histopathological images that are utilized and archived every day. Such WSIs are acquired from glass histology slides using dedicated scanning devices after a staining process. In each WSI, thousands of nuclei from various types of cell can be identified. The detection of such nuclei is crucial for the identification of tissue structures, which can be further analyzed in a systematic manner and used for various clinical tasks. Presence, extent, size, shape, and other morphological characteristics of such structures are important indicators of the severity of different diseases~\cite{gleason1992histologic}. Moreover, a quantitative analysis of digital pathology is important, to understand the underlying biological reasons for diseases~\cite{rubin2008rubin}.

Manual segmentation or estimation of nuclei on a WSI is an extremely time consuming process which suffers from high inter-observer variability~\cite{andrion1995malignant}. On the other hand, data-driven methods that perform well on a specific histopathological datasets report poor performance on other datasets due again to the high variability in acquisition parameters and biological properties of cells in different organs and diseases~\cite{kumar2017dataset}. To deal with this problem, datasets integrating different organs~\cite{kumar2017dataset,gamper2019panuke} based on images from The Cancer Genome Atlas (TCGA) provide pixelwise annotations for nuclei from variety of organs. Yet, these datasets provide access to only a limited range of annotations, making the generalization of these techniques ambiguous and emphasizing the need for  novel segmentation algorithms without relying purely on manual annotations.

To this end, in this paper, we propose a self-supervised approach for nuclei segmentation without requiring annotations. The contributions of this paper are threefold: (i) we propose using scale classification as a self-supervision signal under the assumption that nuclei are a discriminative feature for this task; (ii) we employ a fully convolutional attention network based on dilated filters that generates segmentation maps for nuclei in the image space; and (iii) we investigate regularization constraints on the output of the attention network in order to generate semantically meaningful segmentation maps. 


\section{Related Work}
 Hematoxylin and eosin (H\&E) staining is one of the most common and inexpensive staining schemes for WSI acquisition.
 A number of different tissue structures can be identified in H\&E images such as glands, lumen (ducts within glands), adipose (fat), and stroma (connective tissue). The building blocks of such structures are a number of different cells. During the staining process, hematoxylin renders cell nuclei  dark  blueish  purple  and  the epithelium  light  purple, while eosin renders stroma pink. A variety of standard image analysis methods are based on hematoxylin in order to extract nuclei~\cite{yi2017extraction,boyle2014prognostic} reporting very promising results, albeit evaluated mostly on single organs. A lot of research on the segmentation of nuclei in WSI images has been presented over the past few decades. Methodologies that integrate thresholding, clustering, watershed algorithms, active contours, and variants along with a variety of pre- and post-processing techniques have been extensively studied~\cite{gurcan2009histopathological}. A common problem among the aforementioned algorithmic approaches is the poor generalization across the wide spectrum of tissue morphologies introducing a lot of false positives. 

To counter this, a number of learning-based approaches have been investigated in order to better tackle the variation over nuclei shape and color. One group of learning-based methods includes hand-engineered representations such as filter bank responses, geometric features, texture descriptors or other first order statistics paired with a classification algorithm~\cite{kong2011partitioning,plissiti2012overlapping}. 
Recent success of deep learning-based methods and the introduction of publicly available datasets~\cite{kumar2017dataset,gamper2019panuke} formed a second learning-based group of supervised approaches. In particular,~\cite{kumar2017dataset} summarises some of these supervised approaches that are developed for multi-organ nuclei segmentation, most of them based on convolutional neural networks. Among them the best performing method proposes a multi-task scheme based on an FCN~\cite{long2015fully} architecture using a ResNet~\cite{he2016deep} backbone encoder with one branch to perform nuclei segmentation and a second one for contour segmentation. Yet, the emergence of self-supervised approaches in computer vision~\cite{pathak2017curiosity,gidaris2018unsupervised} has not successfully translated to applications in histopathology. In this paper, we proposed a self-supervised method for nuclei segmentation exploiting magnification level determination as a self-supervision signal.

  \begin{figure}[t!]
   \centering
   \includegraphics[width=\linewidth]{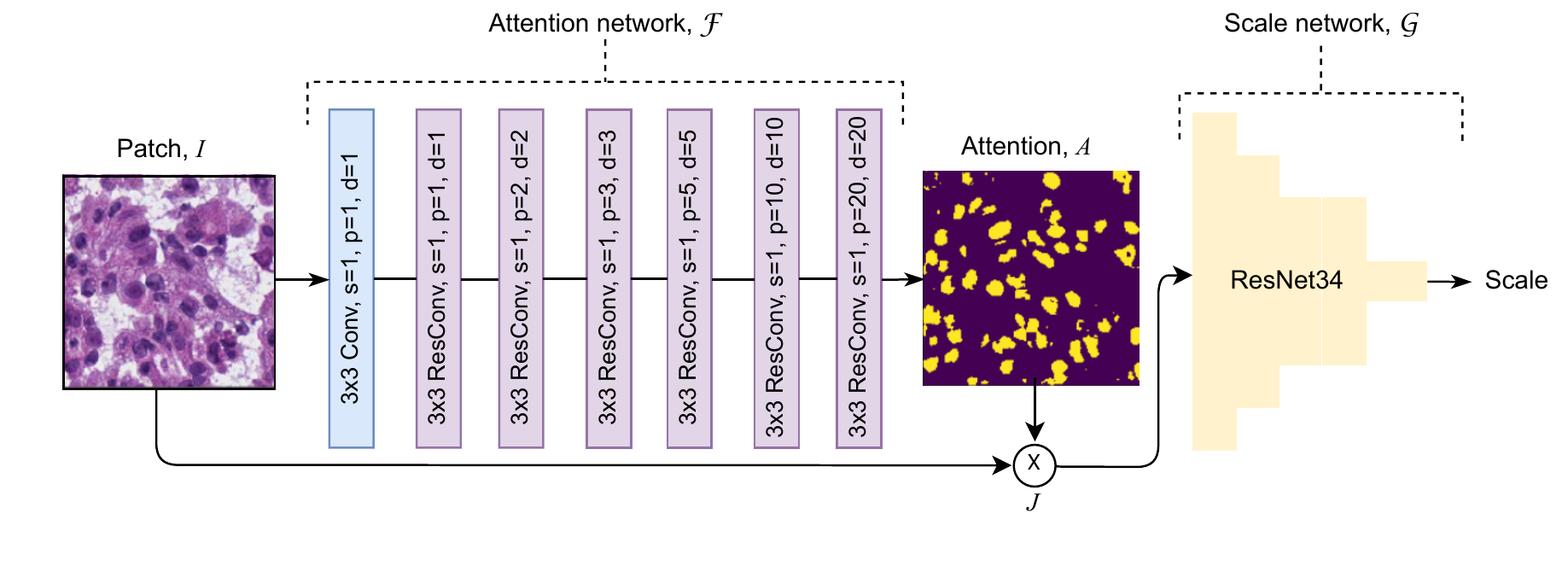}
   \caption{A diagram of our approach. Each patch  $\patch$ is fed to the attention network $\fex$ generating an attention map $\attimg$. The ``attended'' image $\attpatch$ is then given to the scale classification network $\scalenet$. Both networks are trained in an end-to-end fashion. \texttt{s}, \texttt{p}, and \texttt{d} for convolution blocks refer to stride, padding, and dilation.}
   \label{fig:model}
  \end{figure}

 \section{Methodology}
  The main idea behind our approach is that given a patch extracted from a WSI viewed at a certain magnification, the level of magnification can 
  be ascertained by looking at the size and texture of the nuclei in the patch. By extension, 
  we further assume that the nuclei are enough to determine the level of
  magnification, and other artefacts in the image are not necessary for 
  this task. 
  
  Contrary to several concurrent computer vision pipelines which propose to train 
  and evaluate models by feeding images sampled at several scales 
  in order for them to learn multi-scale features~\cite{he2016deep} or models which specifically
  train for scale equivariance~\cite{worrall2019deep}, we posit learning a 
  scale-\emph{sensitive} network which specifically trains for discriminative
  features for correct scale classification\protect\daggerfootnote{Note that the terms \emph{scale} (in the context of computer vision) and \emph{magnification} (in the context of histopathology) are semantically equivalent and used interchangeably.}. Given a set of WSIs, we extract all 
  tissue patches (tiles) from them at a fixed set of 
  magnifications $\levels$. 
  We consider only these tiles, with the ``ground-truth'' knowledge for each tile
  being at what magnification level it was extracted. 
  Following our earlier reasoning, if nuclei in a given tile 
  $\patch \in \mathbb{R}^{3\times H\times W}$ are enough to predict the level of 
  magnification, we assume that there exists a corresponding attention map $\attimg$, so that
  $\attimg \odot \patch$ is also enough to determine the magnification, where 
  $\odot$ represents element-wise multiplication, and 
  $\attimg \in [0,1]^{1\times H\times W}$ is a single channel attention
  image that focuses on the nuclei in the input tile (Figure~\ref{fig:model}).
  
    We design a fully-convolutional feature extractor $\fex$ to predict the attention
  map $\attimg$ from the patch $\patch$. Our feature extractor consists of 
  several layers of convolution operations with a gradual increase in the 
  dilation of the kernels so as to incorporate information from a large 
  neighborhood around every pixel. This feature extractor $\fex$ regresses a 
  confidence map $\attconf = \fex\left(\patch\right) \in \mathbb{R}^{1\times H\times W}\,,$ which is activated by a compressed and biased sigmoid function so that
  $\attimg = \sigma\left(\attconf\right)$. In order to force the attention map 
  to focus only on parts of the input patch, we apply a sparsity regularizer 
  on $\attimg$. This regularizer follows the idea and implementation of a 
  concurrent work on unsupervised separation of nuclei and background~\cite{hou2019sparse}. Sparsity is imposed by picking the $\sparseparam$-th 
  percentile value in the confidence map $\attconf$ for all images
  in the batch, and choosing a threshold $\tau$ equal to the average of this
  percentile over an entire training batch. Formally, 
  \begin{equation}
      \tau = \frac{1}{B}\sum_{b=1}^{B}\attconf_b^{(\sparseparam)}\,,
  \end{equation}
  where $\attconf_b^{(\sparseparam)}$ represents the $\left(\frac{\sparseparam}{100}\cdot HW\right)$-th largest value in the confidence 
  map $\attconf_b$ for the $b$-th image in the training batch of $B$ images. The sigmoid is then defined as $\sigma(x) = \frac{1}{1 + \exp\left(-r\left( x - \tau\right)\right)}\,.$
  It is compressed in order to force sharp transitions in the activated attention
  map, the compression being determined by $r$. We use $r = 20$ in our experiments.

  
  The ``attended'' image $\attpatch = \attimg \odot \patch$ is now enough for 
  magnification or scale classification. We train a scale classification
  network $\scalenet$, which we initialize as a ResNet-34~\cite{he2016deep}, 
  to predict the magnification level for each input tile
  $\attpatch$. The output of this network is scores for each magnification
  level, which is converted to probabilities using a softmax activation. 
  The resulting model (Figure~\ref{fig:model}) is trainable in an
  end-to-end manner. We use negative log-likelihood to train the scale classification
  network $\scalenet$, and in turn the attention network $\fex$---
  \begin{equation}
      \loss{scale}\left(\lprobs, l\right) = -\log\lprob_l\,;\text{~}  
      \lprob_i = \left[\mathrm{softmax}\left(\lscores\right)\right]_i\,;\text{~}
      \lscores = \scalenet\left(J\right)\,;
      \text{~} \,1 \le i \le \nlevels\,,
  \end{equation}
  where $l$ is the scale ground-truth, and $\nlevels = \left|\levels\right|\,.$
   
  \subsection{Smoothness Regularization}
   We wish $\attimg$ to be semantically meaningful and smooth with blobs focusing on nuclei
   instead of having high frequency components. To this end, we incorporate a 
   smoothness regularizer on the attention maps. The smoothness regularizer
   attempts simply to reduce the high frequency component that might appear
   in the attention map because of the compressed sigmoid.
      We employ a standard smoothness regularizer based on spatial gradients defined as 
   \begin{equation}
       \loss{smooth} = \frac{1}{(H-1)(W-1)}\sum_{i,j}\normone{A_{i+1,j}-A_{i,j}} + \normone{A_{i,j+1} - A_{i,j}}\,.
   \end{equation}

  \subsection{Transformation Equivariance}
   Equivariance is a commonly used constraint on feature extractors for imposing
   semantic consistency~\cite{thewlis2017unsupervised,cohen2019gauge}. A feature extractor $f$ is equivariant to a transformation $g$ if $g$ is replicated in the feature vector produced by $f\,,$ i.e., $f(g(\mathbf{x})) = g(f(\mathbf{x}))\,,$ for an image $\mathbf{x}$.  
   In the given context, we want the attention map obtained from $\fex$ to be 
   equivariant to a set $\settransforms$ of certain rigid transforms. We impose equivariance to these transformations through
   a simple mean squared error loss on $\attimg$. Formally, we define
   the equivariance constraint as 
   \begin{equation}
       \loss{equiv} = \frac{1}{HW}\normtwo{\sigma\left(t\left(\fex\left(\patch\right)\right)\right) - \sigma\left(\fex\left(t\left(\patch\right)\right)\right)}^2\,,
   \end{equation}
   for a transformation $t \in \settransforms$. We set $\settransforms$ to include
   horizontal and vertical flips, matrix transpose, and rotations by $90$, 
   $180$, and $270$ degrees.
   Each training batch uses a random $t \in \settransforms\,.$


  \subsection{Training}
   The overall model is trained in an end-to-end fashion, with $\loss{scale}$ being the guiding self-supervision loss. For models incorporating all constraints, i.e., smoothness, sparsity, and equivariance, the total loss is
   \begin{equation}
       \loss{total} = \loss{scale} + \loss{smooth} + \loss{equiv}\,.
   \end{equation}
   We refer to a model trained with all these components together as $\modeltotal$. We also test models without one of these losses to demonstrate how each loss contributes to the learning. More specifically, we define the following models: 
   
   \begin{enumerate}
       \item $\modelNsmooth$: does not include $\loss{smooth}$. 
       \item $\modelNequiv$: does not include $\loss{equiv}$. 
       \item $\modelNsparse$: does not include a sparsity regularizer on the attention map. In this case, the sigmoid is simply defined as 
       $\sigma(x) = \frac{1}{1 + \exp(-x)}\,$. 
       \item $\modelNWSI$: a model which does not sample images from WSIs, but instead from a set of pre-extracted patches (see Section~\ref{sec:dataset}).
   \end{enumerate}
   
   We set the sparsity parameter $\sparseparam$ empirically in order to choose the $93$-rd percentile value for sparsity regularization. This is equivalent to assuming that, on an average, $7\%$ of the pixels in a tile represent nuclei. 

   \subsection{Post Processing, Validation, and Model Selection}
   \label{sec:postprocess}
   In order to retrieve the final instance segmentation from the attention image we employ a post processing pipeline that consists of 3 consequent steps. Firstly, two binary opening and closing morphological operations are sequentially performed using a coarse and a fine circular element ($r=2$, $r=1$). Next, the distance transform is calculated and smoothed using a Gaussian blur ($\sigma=1$) on the new attention image and the local maxima are identified in a circular window ($r=7$). Lastly, a marker driven watershed algorithm is applied using the inverse of the distance transform and the local maxima as markers.
   
   As our model does not explicitly train for segmentation of nuclei, we require a validation set to determine which model is finally best-suited for our objective. To this end, we record the Dice score between the attention map and the ground truth on the validation set (see Section~\ref{sec:dataset}) at intermediate training epochs, and choose the epoch which performs the best. We noticed that, in general, performance increases initially on the validation, but flattens after $\mathtt{\sim}30$ epochs.

\section{Experimental Setup and Results}

\subsection{Dataset}
\label{sec:dataset}
For the purposes of this study we used the MoNuSeg database~\cite{kumar2017dataset}. This dataset contains thirty $1000 \times 1000$ annotated patches extracted from thirty WSIs from different patients suffering from different cancer types from The Cancer Genomic Atlas (TCGA). We downloaded the WSIs corresponding to patients included in the training split and extracted tiles of size $224\times 224$ from three different magnifications, namely $10\times\,$, $20\times\,$, and $40\times\,$. For each extracted tile, we perform a simple thresholding in the HSV color space to determine whether the tile contains tissue or not. Tiles with less then $70\%$ tissue cover are not used. Furthermore, a stain normalization step was performed using the color transfer approach described in~\cite{reinhard2001color}. Finally, a total of $1\,125\,737$ tiles from the three aforementioned scales were selected and paired with the corresponding magnification level. The MoNuSeg train and test splits were employed, while the MoNuSeg train set was further split into training and validation as $19$ and $11$ examples, respectively. The annotations provided by MoNuSeg on the validation set were utilized for determining the four post processing parameters (Section~\ref{sec:postprocess}) and for the final evaluation. For the model $\modelNWSI$, which does not use whole slide images, we use the MoNuSeg patches instead for training, using the same strategy to split training and validation. 
We further evaluated the performance of our model that was trained on the MoNuSeg training set on the TNBC\cite{naylor2018segmentation} and CoNSeP\cite{graham2019hover} datasets.
   
   \subsection{Implementation}
    We use the PyTorch~\cite{paszke2017automatic} library for our code. We use the Adam~\cite{kingma2014adam} optimizer in all our experiments, with an initial learning rate of $0.0002\,$, a weight decay of $0.0001\,$, and $\beta_{1}=0.9\,$. We use a batch size of $32\,$, $100$ minibatches per epoch, and randomly crop patches of size $160 \times 160$ from training images to use as inputs to our models. Furthermore, as there is a high imbalance among the number of tiles for each of the magnification level (images are about $4$ times more in number for a one step increase in the magnification level), we force a per-batch sampling of images that is uniform over the magnification levels, i.e., each training batch is sampled so that images are divided equally over the magnification levels. This is important to prevent learning a biased model.

\begin{figure*}[t!]
    \centering
        \includegraphics[height=2.8cm]{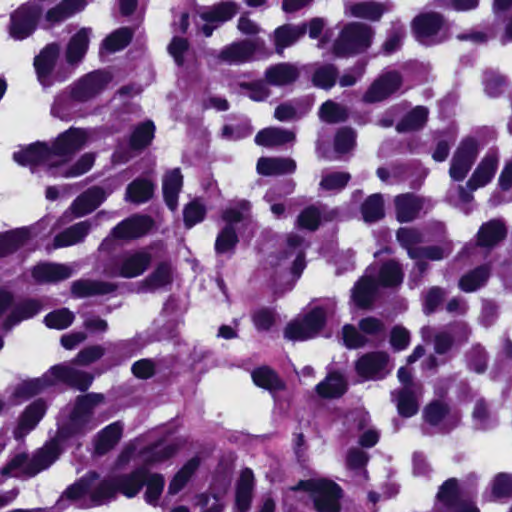}
    ~ 
        \includegraphics[height=2.8cm]{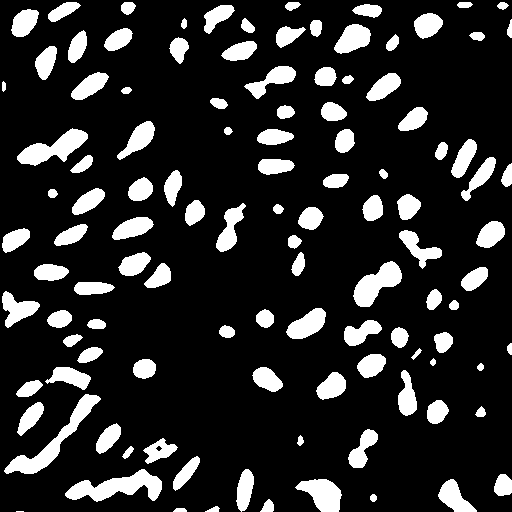}
    ~ 
        \includegraphics[height=2.8cm]{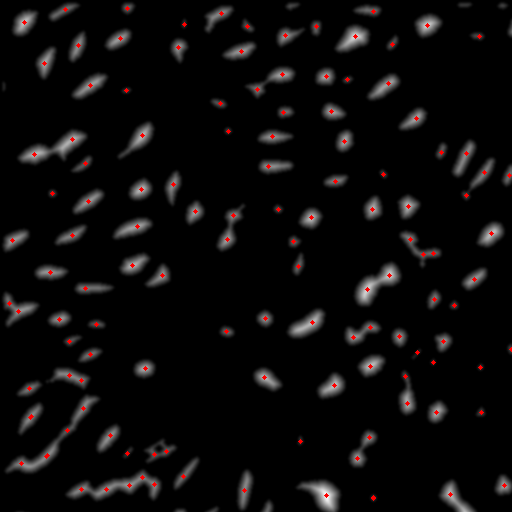}
    ~ 
        \includegraphics[height=2.8cm]{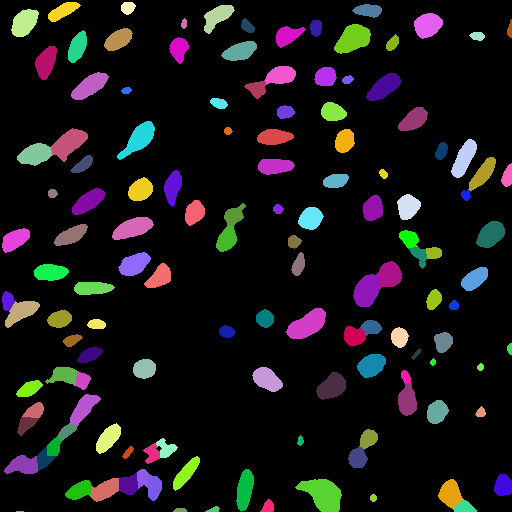}
    \caption{Input, intermediate results and output of the post processing pipeline. From left to right: the input image; the attention map obtained from $\modeltotal$ after the post-processing; the distance transform together with local maxima over-imposed in red; and the final result after the marker driven watershed.} \label{fig:result}
    \label{fig:result}
    \vspace{-15pt}
\end{figure*}







\begin{table}[t!]
\footnotesize
\centering
\begin{tabular}{@{}p{2.4cm} l x{1.6cm} x{1.6cm} x{1.6cm}@{}}
\toprule
                         Test dataset &
                         Method & \begin{tabular}[c]{@{}c@{}}AJI~\cite{kumar2017dataset}\end{tabular} & \begin{tabular}[c]{@{}c@{}}AHD\end{tabular} & \begin{tabular}[c]{@{}c@{}}ADC\end{tabular} \\
\midrule
\multirow{10}{2.4cm}{MoNuSeg test}
& CNN2~\cite{kumar2017dataset}$^\dagger$            & 0.3482                   & 8.6924                    & 0.6928 \\
& CNN3~\cite{kumar2017dataset}$^\dagger$            & 0.5083                   & 7.6615                    & \textbf{0.7623} \\[5pt]
& Best Supervised~\cite{kumar2017dataset}$^\dagger$ & \textbf{0.691}                    & -                         & - \\[5pt]

& CellProfiler~\cite{kumar2017dataset}    & 0.1232                   & 9.2771                    & 0.5974 \\
& Fiji~\cite{kumar2017dataset}            & 0.2733                   & 8.9507                    & 0.6493 \\ [5pt]

& $\modelNsparse$                 & 0.0312                    & 13.1415     & 0.2283 \\
& $\modelNsmooth$               & 0.1929                    & 8.8166                        & 0.4789 \\  
& $\modelNWSI$         & 0.3025 & 8.2853 & 0.6209 \\[5pt]
& $\modelNequiv$             & 0.4938                    & 8.0091                        & 0.7136 \\          
& $\modeltotal$                             & \textbf{0.5354}                &  \textbf{7.7502}                        & \textbf{0.7477} \\ 
\midrule

\multirow{5}{2.4cm}{TNBC\cite{naylor2018segmentation}} 
& U-Net\cite{graham2019hover}$^\dagger$& 0.514 & - & 0.681 \\
& SegNet+WS\cite{graham2019hover}$^\dagger$ & 0.559 & - & \textbf{0.758} \\
& HoverNet\cite{graham2019hover}$^\dagger$& \textbf{0.590} & - & 0.749 \\[5pt]

& CellProfiler & 0.2080 & - & 0.4157 \\[5pt]

& $\modeltotal$ & 0.2656 & - & 0.5139 \\

\midrule

\multirow{5}{2.4cm}{CoNSeP\cite{graham2019hover}}
  & SegNet\cite{graham2019hover}$^\dagger$& 0.194 & - & \textbf{0.796} \\
 & U-Net\cite{graham2019hover}$^\dagger$& \textbf{0.482} & - & 0.724 \\[5pt]
 
 & CellProfiler\cite{graham2019hover} & 0.202 & - & 0.434 \\
 & QuPath\cite{graham2019hover} & 0.249 & - & 0.588 \\[5pt]
 
 & $\modeltotal$ & 0.1980 & - & 0.587 \\
 
\bottomrule

\end{tabular}
\caption{Quantitative results of the different benchmarked methods on three different public available datasets. AJI, AHD, and ADC stand for Aggregated Jaccard Index, Average Hausdorff Distance, and Average Dice Coefficient, respectively. Methods marked with $^\dagger$ are supervised.} \label{tab:res}
\end{table}




\subsection{Results}
To highlight the potentials of our method we compare its performance with supervised and unsupervised methods on the MoNuSeg testset presented in~\cite{kumar2017dataset}. In particular, in Table~\ref{tab:res} we summarize the performance of three supervised methods (CNN2,CNN3 and Best Supervised) and two completely unsupervised methods (Fiji and CellProfiler) together with different variations of our proposed method. 
Our method outperforms the unsupervised methods, and it reports similar performance with CNN2\cite{kumar2017dataset} and CNN3\cite{kumar2017dataset} on the same dataset. While it reports lower performance than the best supervised method from~\cite{kumar2017dataset}, our formulation is quite modular and able to adapt multi-task schemes similar to the one adapted by the winning method of~\cite{kumar2017dataset}. 

On the TNBC and CoNSeP datasets, our method is strongly competitive among the unsupervised methods. We should emphasize that these results have been obtained without retraining on these datasets. The CoNSeP dataset consists mainly of colorectal adenocarcinoma which is under-represented in the training set of MoNuSeg, proving very good generalization of our method.

Moreover, from our ablation study (Table~\ref{tab:res}), it is clear that all components of the proposed model are essential. Sparsity is the most important as by removing it, the network regresses an attention map that is too smooth and not necessarily concentrating on nuclei, thus being semantically meaningless. Qualitatively, we observed that $\loss{smooth}$ allows the network to focus on only on nuclei by removing attention over adjacent tissue regions, while $\loss{equiv}$ further refines the attention maps by imposing geometric symmetry. Finally, in Figure~\ref{fig:result} the segmentation map for one test image is presented. Results obtained from the $\modeltotal$ attention network together with the nuclei segmentation after the performed post-processing are summarised.


\section{Conclusion}
In this paper, we propose and investigate a self-supervised method for nuclei segmentation of multi-organ histopathological images. In particular, we propose the use of the scale classification as a guiding self-supervision signal to train an attention network. We propose regularizers in order to regress attention maps that are semantically meaningful. Promising results comparable with supervised methods tested on the publicly available MoNuSeg dataset indicate the potentials of our method. We show also via. experiments on TNBC and ConSeP that our model generalizes well on new datasets. In the future, we aim to investigate the integration of our results within a treatment selection strategy. Nuclei presence is often a strong bio-marker as it concerns emerging cancer treatments (immunotherapy). Therefore, the end-to-end integration coupling histopathology and treatment outcomes could lead to prognostic tools as it concerns treatment response. Parallelly, other domains in medical imaging share concept similarities with the proposed concept. 
%
%
%
\bibliographystyle{splncs04}
\bibliography{arxiv}
%








\end{document}